\documentclass[aps,prl,reprint,superscriptaddress,nobibnotes]{revtex4-2}

\usepackage{blindtext}
\usepackage{amsthm, amsfonts}
\usepackage{amsmath, amssymb}
\usepackage[normalem]{ulem}
\usepackage{xcolor}
\usepackage{graphicx}
\usepackage{mathtools}
\usepackage{physics}
\usepackage{hyperref}

\newcommand{\dup}{\text{d}}
\newcommand{\1}{\text{\usefont{U}{bbold}{m}{n}1}}
\newcommand{\tp}{\otimes}
\newcommand{\R}{\mathbb{R}}

\begin{document}

\title{Discovering Local Hidden-Variable Models for Arbitrary Multipartite Entangled States and Arbitrary Measurements}

\author{Nick von Selzam}

\email{nick.von-selzam@mpl.mpg.de}

\affiliation{Max Planck Institute for the Science of Light, 91058 Erlangen, Germany}

\author{Florian Marquardt}

\affiliation{Max Planck Institute for the Science of Light, 91058 Erlangen, Germany}
\affiliation{Department of Physics, Friedrich-Alexander-Universität Erlangen-Nürnberg, 91058 Erlangen, Germany\\}

\date{\today}

\begin{abstract}

Measurement correlations in quantum systems can exhibit non-local behavior, a fundamental aspect of quantum mechanics with applications such as device-independent quantum information processing. However, the explicit  construction of local hidden-variable (LHV) models remains an outstanding challenge in the general setting. To address this, we develop an approach that employs gradient-descent algorithms from machine learning to find LHV models which reproduce the statistics of arbitrary measurements for quantum many-body states. In contrast to previous approaches, our method employs a general ansatz, enabling it to discover an LHV model in all cases where the state is local. Therefore, it provides actual estimates for the critical noise levels at which two-qubit Werner states and three-qubit GHZ and W states become non-local. Furthermore, we find evidence suggesting that two-spin subsystems in the ground states of translationally invariant Hamiltonians are local, while bigger subsystems are in general not. Our method now offers a quantitative tool for determining the regimes of non-locality in any given physical context, including scenarios involving non-equilibrium and decoherence. 

\end{abstract}

\maketitle

\section{Introduction}

Bell famously showed that quantum mechanics predicts states which display non-local measurement correlations \cite{Bell64}. By now this has also been confirmed experimentally \cite{CHSH69, Aspect82, Weihs98, Christensen13, Hensen15, Giustina15, Shalm15, Rosenfeld17}. It remains a difficult task to determine whether any given state is local or not, though. Even for the simple example of two-qubit Werner states \cite{Werner89} -- mixtures of the Bell singlet with white noise -- the critical visibility at which these switch from being local to non-local is not known. While entanglement is necessary for non-locality \cite{Werner89} and pure states are non-local if and only if they are entangled \cite{Popescu92, Gisin92}, there are entangled mixed states which nevertheless remain local \cite{Werner89}. That is, entanglement and non-locality are inequivalent resources \cite{Augusiak15, Bowles16}. 

In this work, we present a general approach to solve the challenge of determining whether a given quantum state is local. Specifically, we use machine learning methods to discover local hidden-variable (LHV) models which reproduce the measurement statistics of quantum many-body states.

For fixed finite measurement options one would obtain a finite set of correlations. Locality of these correlations is fairly well understood and there exist efficient algorithms which determine whether such correlations are local or not (see e.g. \cite{Brunner14} and references therein). However, to reliably decide whether a given state is local, we are interested in the significantly more general case of an infinite continuum of measurement options (e.g., all projective measurements). We demonstrate locality by explicitly constructing local models that take into account this whole continuum. Only checking particular Bell inequalities, which is often done in studies of non-locality in quantum many-body systems, is not sufficient to solve this challenge.

To demonstrate the broad applicability of our approach we illustrate it within two different research areas. 
From the quantum information perspective the noise robustness of non-locality is of interest. Importantly, there is a series of works establishing bounds on the critical visibility of the two-qubit Werner states \cite{CHSH69, Werner89, Acin06, Vertesi08, Hua15, Brierley16, Divianszky17, Hirsch17, Designolle23}. With our method we significantly improve on this by providing an actual estimate for the critical visibility. Likewise, we are able to obtain estimates for the critical visibilities of noisy three-qubit GHZ and W states. 

From the condensed matter perspective the study of non-locality in quantum many-body systems has long been of interest \cite{Batle10, Justino12, CM1, CM2, CM3, CM4, CM5, CM6, CM7, CM8, CM9, CM10, CM11} (see \cite{Chiara18} for a review). It has been demonstrated that for some systems simple Bell inequalities, such as the CHSH inequality, are not violated in the ground state or even during non-equilibrium time evolution \cite{Batle10, Justino12}. In fact, there exists a \lq monogamy of Bell correlations\rq\ relation \cite{Toner06} which implies that the CHSH Bell-inequality is never violated in translation-invariant systems \cite{Oliveira13, Sun13}. We can now take the same kind of systems and analyze whether these states are also genuinely local. Moreover, we are able to go beyond two-spin subsystems and study locality of larger subsystems for which even less is known.

We mention some related approaches: 
In \cite{Deng18} restricted Boltzmann machines and reinforcement learning are used to find maximal violations of a given Bell inequality. In \cite{Canabarro19} neural networks together with genetic algorithms are used to predict a non-locality quantifier for given finite correlations. In \cite{Krivachy20} neural networks are used to construct LHV models in network scenarios for finite measurement settings.
In \cite{daSilva23} a similar task is addressed without neural networks.
In \cite{Cavalcanti16, Hirsch16, Designolle23} locality of states is analyzed using a combination of numerical and analytic methods. There are also some purely analytic results \cite{Toner07}. To the best of our knowledge all these approaches have at least one of the following shortcomings: They only allow for a finite set of measurement settings, they only consider a single Bell inequality or they are specialized to specific states. Additionally, they are often not constructive. With our numerical scheme we are able to overcome all of these restrictions simultaneously.

\section{Physical LHV Models}
\label{sec: physical LHV models}

An LHV model describes a particular form of joint measurement statistics. For~$N$ parties the probabilities for obtaining measurement results~$a = (a_1,\ldots,a_N)$ given measurement settings~$x = (x_1,\ldots,x_N)$ are (Bell-)local if they can be written as \cite{Bell64}
\begin{align}
    \label{eq: LHV correlations standard}
    P^{LHV}(a\, |\, x) = \int_\Lambda \dup \lambda\, p(\lambda)\prod_{j=1}^N q_j(a_j | x_j, \lambda).
\end{align}
Here,~$\lambda\in \Lambda$ is the hidden variable with some distribution~$p(\lambda)$. The \emph{measurement rule}~$q_j(a_j|x_j,\lambda)$ is the probability for the~$j^{\text{th}}$ party to measure~$a_j$ given their input~$x_j$ and a fixed value of the hidden variable. 

\begin{figure}[!ht]
    \centering
    \includegraphics[width=0.45\textwidth]{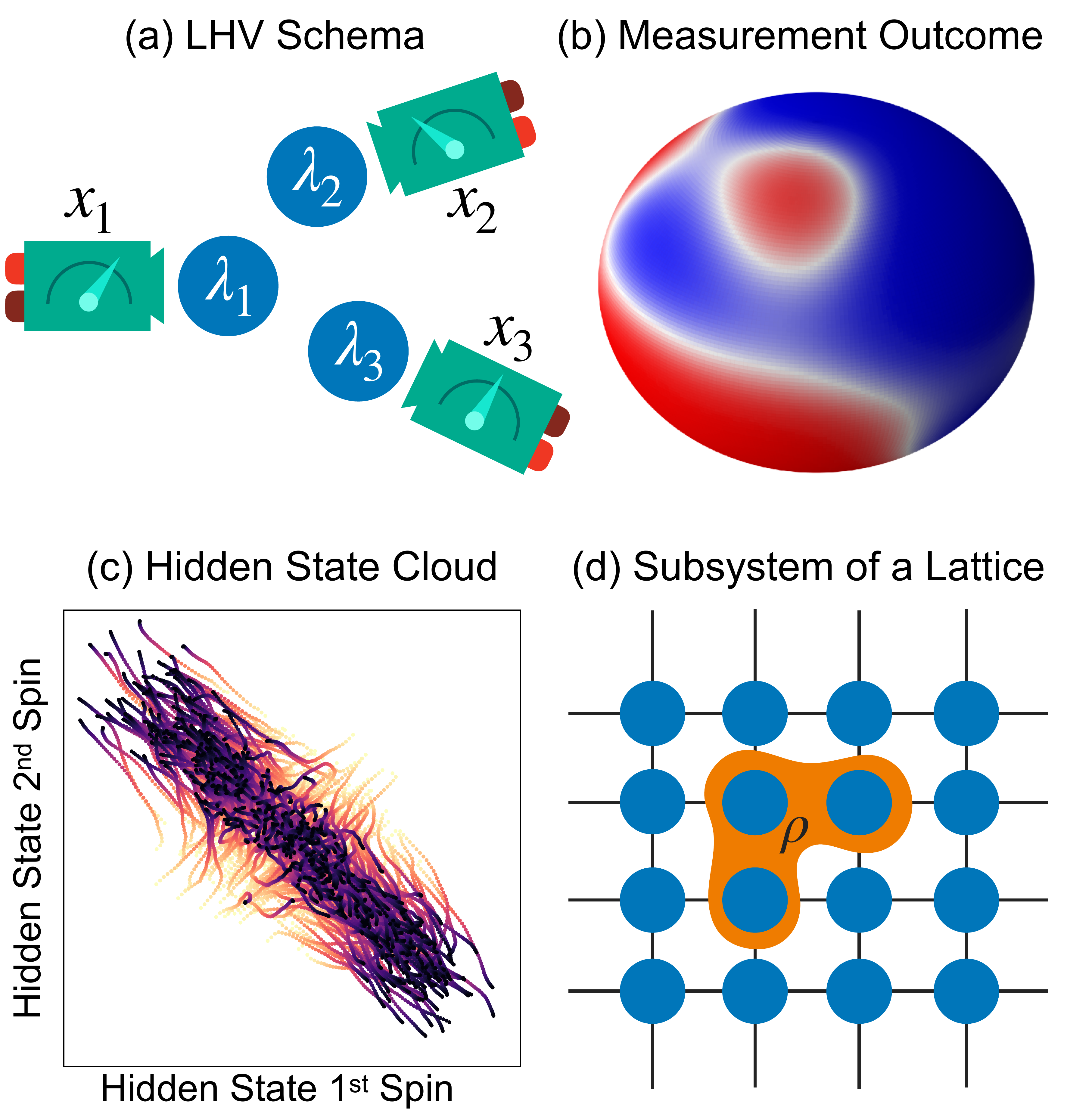}
    \caption{(a) Schematic representation of the LHV setup: Each particle is described by a local hidden-state vector~$\lambda_j$ which determines the measurement outcomes (here binary) depending on the continuous inputs~$x_j$.
    (b): The local measurement outcome -- in the case of spin-$1/2$ systems the probability to measure \lq up\rq\ along all possible directions~$\hat n\in S^2$ -- for a fixed hidden state~$\lambda$. Color coding: Red for~$q(\uparrow|\hat n, \lambda)=1$ (result \lq up\rq) and blue for~$q(\uparrow|\hat n, \lambda)=0$ (result \lq down\rq) interpolating via white. 
    (c): Evolution of the hidden-state cloud (representing the hidden-state distribution~$p(\lambda)$) for two spins during the optimization procedure. Each point corresponds to a pair of hidden states. The color-coding represents the gradient descent step: The initial uncorrelated cloud is yellow, the final (anti-)correlated one is black.
    (d): Subsystems of lattice ground states (translation invariant states). These are generically not too strongly entangled mixed states. We expect them to often be local. }
    \label{fig: 1}
\end{figure}

Our goal is to match $P^{LHV}$ to the quantum measurement statistics of an arbitrary~$N$-particle state~$\rho$. In that scenario, we may reformulate (\ref{eq: LHV correlations standard}) in an equivalent but more physical way, where each particle carries its own hidden variable. As we will show below, this allows us to set up a description where the measurement rule is independent of the underlying quantum state $\rho$, as would be plausibly expected. Mathematically, each particle~$j$ is now described by a local hidden state~$\lambda_j$ as indicated in Fig.~\ref{fig: 1} (a). Upon measurement the measurement rule~$q_j$ describes the outcome as visualized in Fig.~\ref{fig: 1} (b). Information about the state~$\rho$ is only encoded in the \emph{hidden-state distribution}~$p(\lambda_1,\ldots,\lambda_N)$.

Typically, the particles will be of the same type. In this case the measurement in- and outputs are all of the same form. For example, in the case of spin-$1/2$ particles the measurement choice is a normal vector -- e.g., the orientation of a Stern-Gerlach magnet -- and the possible results are \lq up\rq\ or \lq down\rq. This is the same for every spin. Therefore, the hidden states are now of the same form as well,~$\lambda_j\in\Lambda$. Moreover, the measurement rule does not depend on the particle anymore -- it makes no difference whether we measure the first or the second spin. We have the following equivalent definition of local correlations which incorporates all of this, with~$\lambda=(\lambda_1,\ldots,\lambda_N)$
\begin{align}
    \label{eq: LHV correlations physical}
    P^{LHV}(a \, |\, x) = \int_{\Lambda^N}\dup \lambda\, p(\lambda) \prod_{j=1}^N q(a_j | x_j, \lambda_j)
\end{align}
(see Appendix F). 

In the most general case the inputs~$x_j$ correspond to arbitrary POVM measurements. The quantum mechanical measurement probabilities depend only on the outcome (they are non-contextual): Suppose the measurement operator representing the outcome for particle~$j$ is~$a_j$, $a = (a_1,\ldots,a_N)$ (for projective measurements these would be projection operators). Then
\begin{align}
    \label{eq: QM measurement statistics}
    P^{QM}_\rho(a|x) = \Tr(a_1\tp\cdots \tp a_N \cdot \rho).
\end{align}
For all possible measurements the LHV prediction should match the quantum mechanical one. It follows that we may restrict ourselves to non-contextual measurement rules (see Appendix G). Once we fix a sufficiently general measurement rule it is an optimization task to find the best hidden-state distribution given any quantum state
\begin{align}
    \rho \xmapsto{?} p(\lambda).
\end{align}
Note that there are hidden-state distributions not corresponding to any quantum state (see Appendix E).

\section{Optimization}
\label{sec: optimizing the hidden state distribution}

We have set up the structure of our LHV models. Now given an $N$-particle state~$\rho$, how do we optimize the hidden-state distribution~$p(\lambda_1,\ldots,\lambda_N)$ such that the resulting measurement statistics (eq.~(\ref{eq: LHV correlations physical})) are as close as possible to the quantum mechanical ones (eq.~(\ref{eq: QM measurement statistics}))? Borrowing some machine learning terminology, we achieve this by minimising a scalar loss function: a distance measure between~$P^{QM}$ and~$P^{LHV}$. This includes the whole continuum of possible measurements~$x$. Since the method should not be tailored to specific states, all measurement choices are equally important. Therefore, we propose to choose a measure of distance~$\text{D}$ between~$P^{QM}$ and~$P^{LHV}$ for fixed measurement choices and then average over all of them. The resulting deviation between LHV model and QM is of the form
\begin{align}
    \label{eq: loss function}
    \mathcal{L}({LHV}\, ||\,{QM}) = \left\langle \text{D}\hspace{-2pt}\left(P^{QM}(\cdot|x), P^{LHV}(\cdot|x)\right)\right\rangle_{x}.
\end{align}
Since we are dealing with probabilities it is natural to choose the Kullback-Leibler divergence for the distance measure~$\text{D}$.
Note that there are many alternative choices for the loss function. If one is interested only in specific states or measurements, one may tailor it accordingly. In the case of local states a corresponding LHV model leads to a vanishing loss independent of the specific form. However, for non-local states only approximate LHV models can be found and the necessary compromises in the hidden-state distribution will depend on the chosen loss function.

What are the parameters we are optimizing? 
The hidden-state distribution~$p(\lambda_1,\ldots,\lambda_N)$ for~$N$ particles and~$d$-dimensional hidden states ($\lambda_j \in \Lambda = \R^d$) is a function on~$\R^{N\times d}$.
It is infeasible to discretize this space for large~$N$ or~$d$. Therefore, we adopt a Monte-Carlo like approach: The distribution~$p$ is represented by~$N_{\text{h}}$ samples of hidden-state tuples~$(\lambda_1,\ldots,\lambda_N)$.
To minimize the loss function we vary the~$N_\text{h}\cdot N \cdot d$ real parameters of this \emph{hidden-state cloud} (see Fig.~\ref{fig: 1} (c) for a visualization). Note that the average over all measurement settings $x$ in the loss function corresponds to an integral that we can not evaluate. As is common in the field of machine learning we approximate the exact average by repeatedly sampling finite batches of measurement combinations. The optimization is carried out via stochastic gradient descent. 

Optimizing a particle cloud and sampling batches of arbitrarily chosen measurements are the two crucial technical innovations which allow us to go to $N$-particle states and continuous measurements.

We note that such methods -- requiring efficient automatic differentiation with respect to millions of parameters -- have only recently become routinely available through frameworks such as TensorFlow, PyTorch, JAX and due to powerful GPUs.

\section{Spin-1/2 Systems}
\label{sec: spin 1/2 systems}

So far, the setup has been very general. We now make it more concrete by applying it to projective measurements of spin-$1/2$ systems, where we illustrate our approach in several examples. In this case the measurements correspond to asking whether the spin is \lq up\rq\ or \lq down\rq\ (possible measurement results~$a$) along a direction~$\hat n\in S^2$ (measurement choice $x$). 
The average over measurement settings in the loss function (eq.~(\ref{eq: loss function})) becomes an average over directions~$(\hat n_1,\ldots,\hat n_N) \in (S^2)^N$. We simply sample each direction uniformly and independently from the surface of a sphere.  See Appendix H for the specific slightly simplified loss function we use in practice.

\begin{figure*} 
    \centering
    \includegraphics[width=0.99\textwidth]{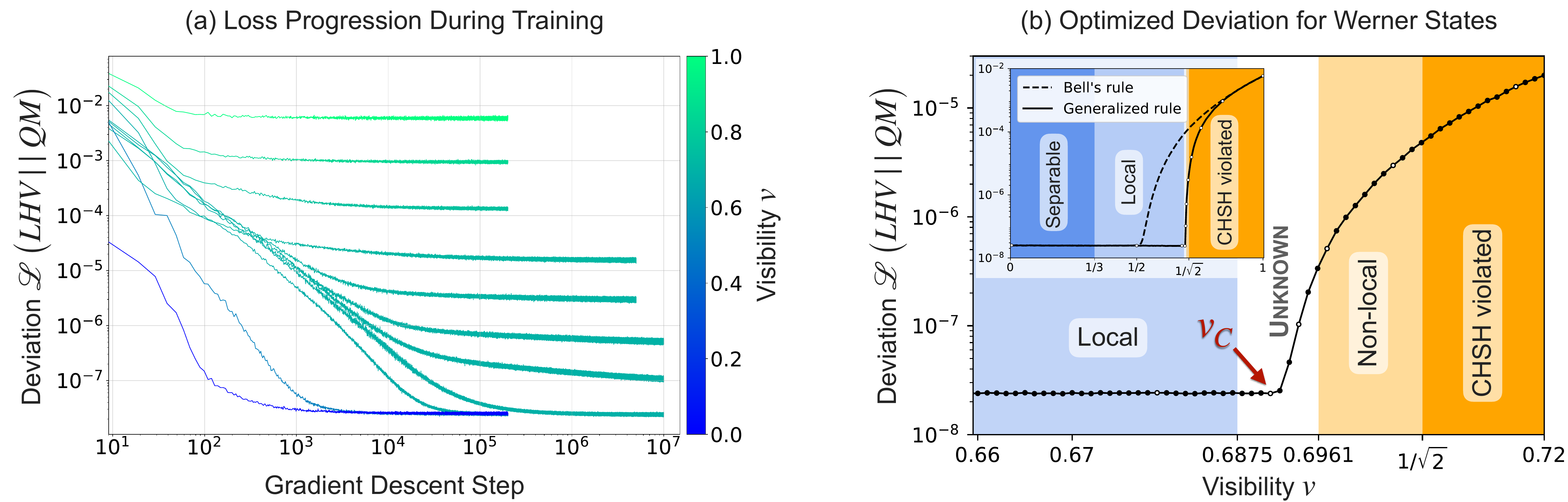}
    \caption{
        \textbf{Optimizing LHV models for two-qubit Werner states.} 
        (a): Deviation~$\mathcal{L}(LHV||QM)$ during training for Werner states with selected visibilities (marked by white dots in (b)). The convergence time depends on the visibility -- becoming larger close to the critical visibility. As demonstrated in Appendix B the optimal hidden-state distribution depends on the visibility in a non-continuous way here, indicating a glassy loss landscape. 
        (b): Deviation~$\mathcal{L}(LHV||QM)$ for the Werner states around the critical visibility (black dots, the lines are guides for the eye). The regions corresponding to local (blue) and non-local (orange) states are highlighted. A subset of the local states are the separable ones (dark blue), a subset of the non-local states are those violating the CHSH inequality (dark orange). In the white gap in-between it has not been known whether the states are local or not. We use the generalized measurement rule with odd spherical harmonics up to degree five to construct the LHV models. 
        Inset: Full range of visibilities (full line) and comparison to Bell's rule (dashed line). The hidden-state cloud size ranges from~$N_{\text{h}}=2^{14}$ to~$N_{\text{h}}=2^{16}$ close to the critical visibility. See Appendix A for details on the~$N_{\text{h}}$-dependence. We observe a small constant loss for all visibilities corresponding to local Werner states and even into the white region. This yields a numerical estimate for the critical visibility:~$v_C\approx 0.691$. With Bell's rule the optimized deviation increases already for~$v\gtrsim 1/2$. It is not expressive enough for these states.}
    \label{fig: Werner}
\end{figure*}    
We now describe how to systematically obtain the most general measurement rule for spin-$1/2$ systems. Due to this generality there is no need for a dependence on the state~$\rho$. 
With a hidden-state space~$\Lambda$, the measurement rule becomes a function~$q: S^2 \times \Lambda \rightarrow [0,\, 1]$. It describes the probability for measuring \lq up\rq\ along the direction~$\hat n$ given a hidden state~$\lambda$: $q({\hat n},\lambda)\equiv q(\lq \rm \text{up} \rq | x={\hat n}, \lambda)$.
It is known that deterministic rules,~$q\in\{0,1\}$, already capture everything (see e.g. \cite{Brunner14}). Therefore, the most general rule can be written as
\begin{align}
    q(\hat n,\lambda) = H(f_{\lambda}(\hat n)),
\end{align}
where~$H$ is the Heaviside step function and the~$f_{\lambda}$ are arbitrary real-valued functions defined on the surface of a sphere. Without loss of generality, we can impose (see Appendix G) 
\begin{align}
    \label{eq: spin measurement rule constraint}
    q(-\hat n,\lambda) = 1-q(\hat n, \lambda).
\end{align}
That is, the~$f_{\lambda}$ are odd functions. We may expand them into (real) odd spherical harmonics (these were also used in a similar but more handcrafted way in \cite{Toner07} for a specific class of states) . Without loss of generality, the expansion coefficients can be taken to become the components of the hidden-state vector~$\vec \lambda\in\R^{\mathbb{N}}$. If~$\vec S(\hat n)$ is a vector containing all odd spherical harmonics in~$\hat n\in S^2$, this can formally be written as
\begin{align}
    q(\hat n,\vec \lambda) = H(\vec S(\hat n)\cdot \vec \lambda).
\end{align}
See Appendix I for a generalization to POVM measurements.

For the numerical implementation we modify two things: First, we introduce a cutoff in the spherical harmonics expansion, writing~$\vec S_D(\hat n)$ for a vector containing all odd spherical harmonics up to degree~$D$. For example (up to normalization factors)
\begin{align}
    \vec S_3(\hat n) \sim (x, y, z, x y z, \ldots, 3 x^2 y - y^3),\quad (x, y, z)=\hat n.
\end{align}
This leads to finite-dimensional hidden variables~$\vec\lambda\in \R^d$, with~$d = \tfrac{1}{2}(D+1)(D+2)$. 
Second, for the numerical optimization we require non-vanishing gradients with respect to the hidden-state vectors. To achieve this, we replace the Heaviside function by a sigmoid function, bringing us back to probabilistic measurement rules:
\begin{align}
    q(\hat n,\vec \lambda) = \sigma(\vec S_D(\hat n)\cdot \vec \lambda),\qquad \sigma(x) = \frac{1}{1+e^{-x}}.
\end{align}
For large norms~$\norm*{\vec\lambda}\to\infty$ we recover the deterministic case.
The choice~$D=1$ would lead to
\begin{align}
    q(\hat n,\hat \lambda) = H(\hat n \cdot \hat \lambda),\qquad \hat\lambda \in S^2,
\end{align}
which was already considered by Bell in his original paper \cite{Bell64}. In Appendix J we show that this \lq Bell's rule\rq at least can represent all separable~$N$-spin-$1/2$ states. 
In all of our examples, including complex many-body states, we find that~$D=5$ already yields results of high accuracy.

We now apply our framework to illustrative examples of states,  both for elementary entangled states and for ground states of strongly correlated quantum many-body systems.

\section{Werner States}
\label{sec: Werner States}

Werner states were introduced to show the existence of entangled states which are local \cite{Werner89}. For two spins they are mixtures of the Bell singlet with the maximally mixed state
\begin{align}
    \rho_v = v\ketbra{\psi} + (1-v) \frac{\1}{4},\quad \ket{\psi} = \frac{1}{\sqrt{2}}\left( \ket{\uparrow \downarrow} - \ket{\downarrow \uparrow} \right).
\end{align}
The parameter~$v\in [0,\, 1]$ is the visibility of the singlet. It is known that these states are separable for~$v\le \frac{1}{3}$ and entangled otherwise. Werner showed that~$\rho_{\frac{1}{2}}$ is still a local state for all projective measurements. Up to the present day, the critical visibility~$v_c$ at which the Werner states cease to be local is unknown. The best bounds so far were obtained in \cite{Designolle23}
\begin{align}
    0.6875 \le v_C \le 0.6961.
\end{align}
Note that the CHSH inequality is only violated for~$v > \frac{1}{\sqrt{2}} \approx 0.7071$, giving a comparatively loose upper bound.

We now apply our algorithm. That is, we optimize hidden-state distributions for Werner states with different visibilities, once with Bell's measurement rule ($D$=1) and once with the general spherical-harmonics rule up to degree five ($D=5$). 
We monitor the loss during the optimization procedure to ensure convergence (Fig.~\ref{fig: Werner}, (a)). The optimized deviations as a function of the visibility are shown in Fig.~\ref{fig: Werner} (b). While Bell's rule is only expressive enough to represent Werner states up to~$v\approx 1/2$, the more general rule can represent all Werner states which are already known to be local. Moreover, it is able to find LHV models  beyond the previously known range, up to a visibility of
\begin{align}
    v_C\approx 0.691. 
\end{align}
This, for the first time, provides an actual numerical estimate for the critical visibility, as opposed to merely lower and upper bounds. In Appendix C we also obtain estimates for the critical visibilities of noisy three-qubit GHZ and W states. 

In order to gauge the precision of these results, we have performed the following additional steps (see Appendix A):
We explain the particular non-zero value of the loss in the local region. For single spin states as well as planar measurements of Werner states we validate that our method agrees with the analytic results. And finally, we observe convergence of the results (e.g., the estimate for the critical visibility) with increasing hidden-state cloud size~$N_{\text{h}}$ and increasing maximal degree~$D$ in the spherical-harmonics expansion. 

We note that the algorithm itself can likely be improved. In particular, there are more sophisticated ways to represent high-dimensional probability distributions such as the hidden-state distribution, for example via generative neural networks. In Appendix B we give details on the complexity and run-time of the algorithm in terms of the hyper parameters.

\section{XXZ Models}
\label{sec: xxz models}

We now turn to quantum many-body systems. In this setting even less is known about locality and researchers have studied mostly the violation of individual Bell inequalities. Local models for~$N$-spin states have been constructed only in very special cases (see e.g. \cite{Toth06} or \cite{Bowles16}).

\begin{figure}[!ht]
    \centering
    \includegraphics[width=0.48\textwidth]{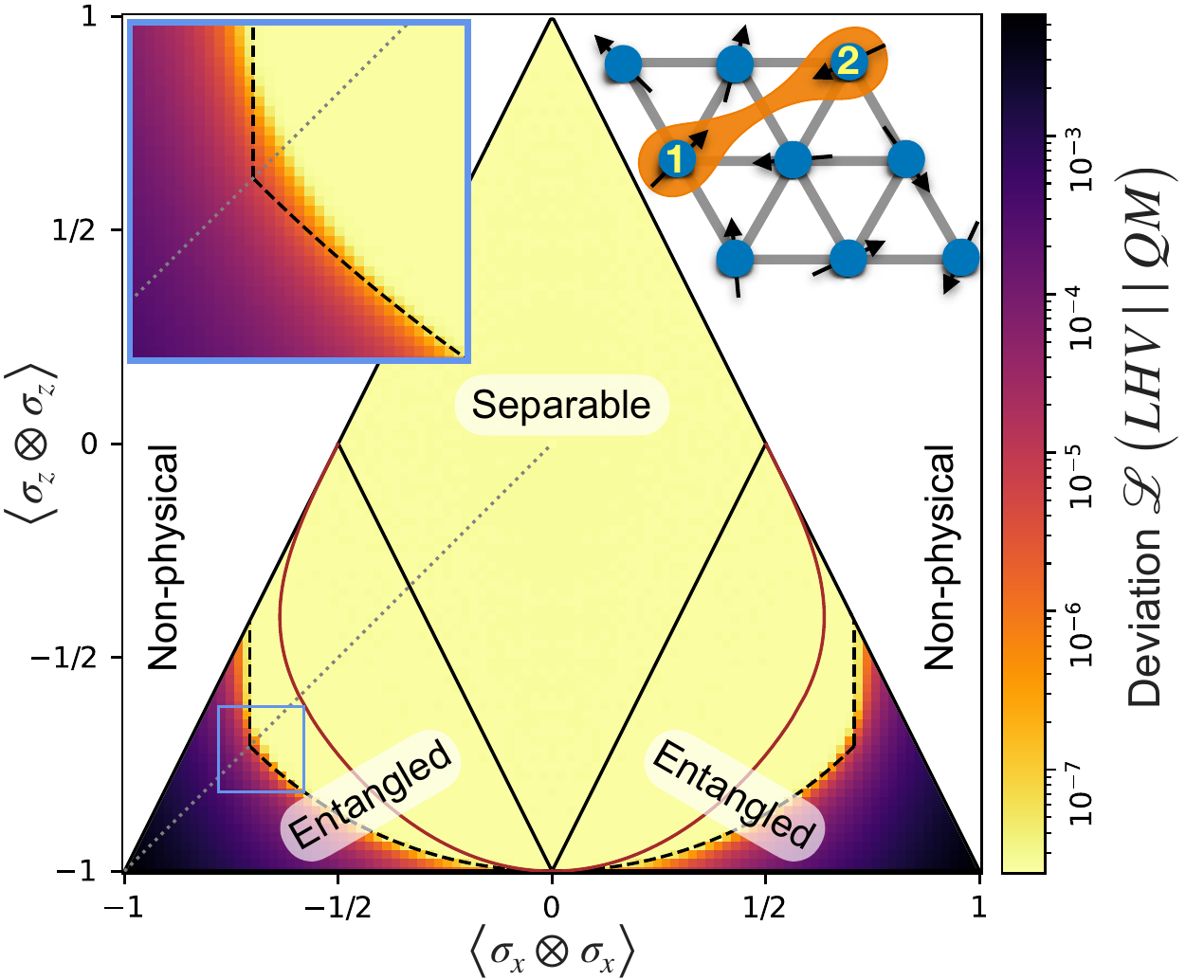}
    \caption{
        \textbf{Locality of two-spin states with the symmetries of the XXZ Hamiltonian.} 
        Deviation~$\mathcal{L}(LHV||QM)$ for all allowed two-spin subsystem states respecting the symmetries of the XXZ Hamiltonian for arbitrary lattice geometries in arbitrary dimensions. The parameter space is divided into non-physical (white), separable (central diamond) and entangled (smaller two triangles) regions. Below the dashed line the CHSH inequality is violated. The red line indicates the boundary of the convex set of density matrices in one-dimensional lattices, that is only the states above the red line can be written as partial traces of translation-invariant states \cite{Verstraete06}. In higher dimensions the corresponding regions are subsets of this. The grey dotted line corresponds to the Werner states. The inset shows a higher-resolution version of the region marked by the blue square. Similar to the Werner states we observe the same small constant value of the loss for all separable and many entangled states. The CHSH inequality is almost tight in this case: For almost all states not violating it we can find local models. Only close to the diagonals (that is, close to the Werner states) this is not the case (see inset). This shows that all physically allowed states in the 1D XXZ model (above the red line) are genuinely local. We show the deviation along the two relevant boundaries (red and dashed lines) in Appendix D . 
        }
    \label{fig: XXZ 2D}
\end{figure}

As an example we analyze locality in the XXZ model, first for two-spin subsystems and then for~$N$-spin subsystems up to~$N=6$. The Hamiltonian is
\begin{align}
    H_{XXZ} = \sum_{\langle j, k\rangle} \left(S^x_j S^x_k + S^y_j S^y_k + \Delta S^z_j S^z_k\right),
\end{align}
where~$S_j^\alpha$ is the spin operator in direction~$\alpha\in\{x, y, z\}$ at site~$j$ and~$\Delta$ is the anisotropy parameter. 

\begin{figure}[!ht]
    \centering
    \includegraphics[width=0.48\textwidth]{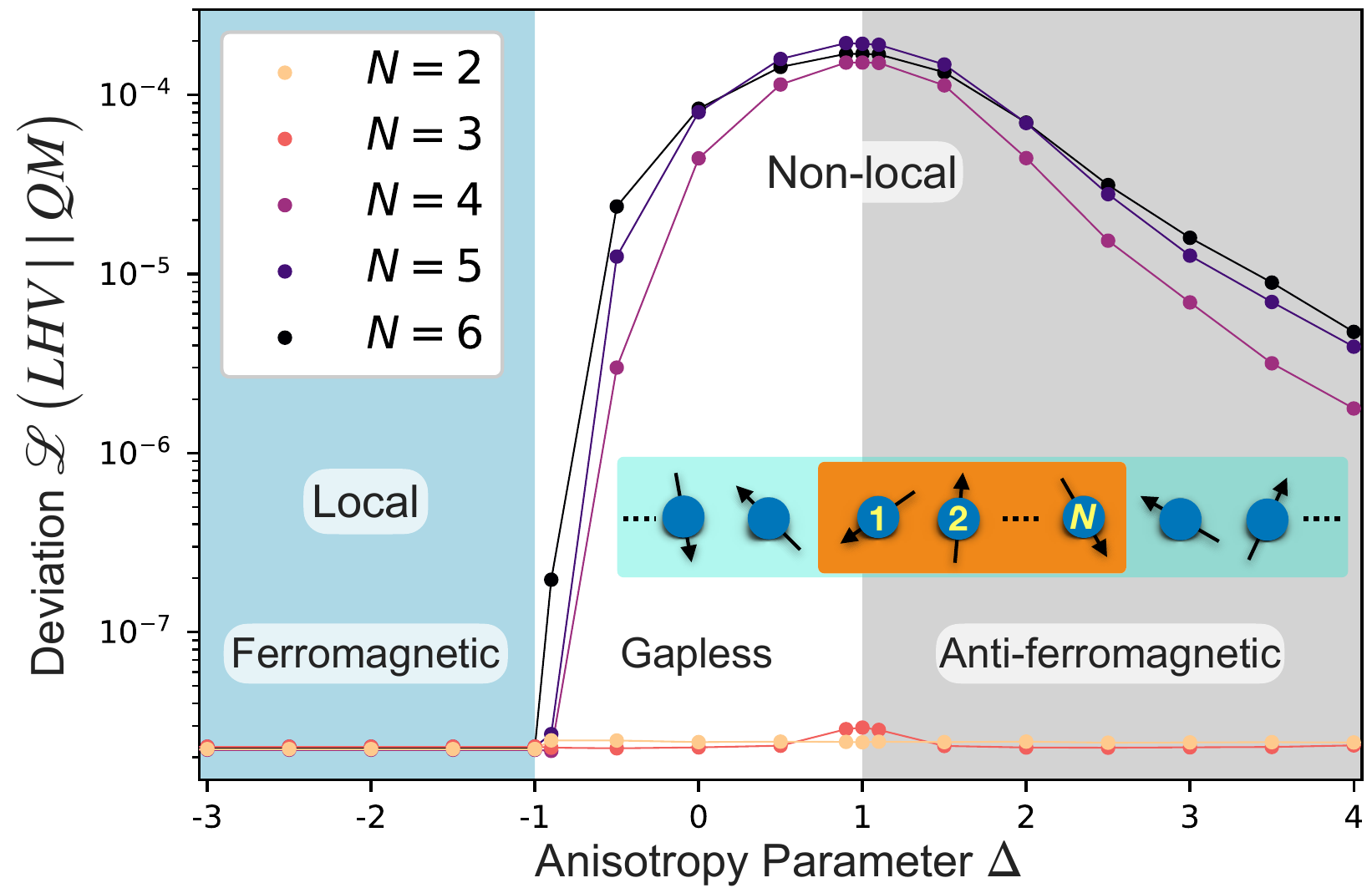}
    \caption{
        \textbf{Locality of~$N$-spin sub-chains in the ground state of a one-dimensional XXZ lattice.} 
        Deviation~$\mathcal{L}(LHV||QM)$ for~$N$-spin sub-chains of a one dimensional XXZ chain with~$22$ spins in total (periodic boundary conditions) as a function of the anisotropy parameter~$\Delta$. The lines are guides for the eye. In the thermodynamic limit they system is in the ferromagnetic phase for~$\Delta < -1$, in the gapless phase for~$-1<\Delta < 1$ and in the anti-ferromagnetic phase for~$\Delta > 1$. The ferromagnetic phase has a product ground state, meaning the correlations in subsystems of any size are local. In the gapless and anti-ferromagnetic phases the ground state is entangled. Two-spin sub-chains remain local though as demonstrated before. Three-spin sub-chains also remain local; only close to the phase transition at~$\Delta = 1$ the optimized deviation slightly increases. Here it is difficult to say whether the corresponding state is genuinely non-local or not, but it is certainly approximately local. For sub-chains of length~$N\ge 4$ the deviation becomes orders of magnitude larger, indicating non-local correlations. Only in the limit~$\Delta\to\infty$ the ground state again becomes a product state and therefore local.}
    \label{fig: XXZ 1D}
\end{figure}

A ground state respects the symmetries of the Hamiltonian (except in cases of degeneracies).
This allows to parameterize all its possible two-spin subsystem states by the correlators~$\langle\sigma_x\tp\sigma_x\rangle$ and~$\langle\sigma_z\tp\sigma_z\rangle$  \cite{Verstraete06}. As a function of these two parameters it is known which correlations correspond to separable or entangled states \cite{Verstraete06}. We also know for which parameters the corresponding state violates the CHSH Bell-inequality \cite{Justino12}.
However, it is an open problem to decide which ones are local and which ones are not. We optimize the hidden-state distribution to find LHV models for the whole range of parameters. The optimized deviations are shown in Figure~\ref{fig: XXZ 2D}.
We find that the CHSH Bell-inequality is almost tight in this case: Most states not violating it can be represented by an LHV model. In particular, all states which are actually realized by infinite XXZ lattice models (that is, those which not only have the correct symmetries but are also partial traces of infinite translation-invariant states) are local. In Appendix D we provide more support for this claim.

Our numerical approach allows us to construct LHV models for more than two spins, which is crucial to acquire a complete understanding of the locality properties of quantum many-body states. To study this regime in an example, we numerically obtain the ground state of a finite one-dimensional XXZ chain with~$22$~spins and optimize LHV models for sub-chains of~$N=2,\ldots,6$ neighboring spins. This is shown in Fig.~\ref{fig: XXZ 1D}. The two-spin subsystems are local for all choices of the anisotropy parameter~$\Delta$, consistent with the results above. The three-spin subsystems are local for almost all~$\Delta$. For~$N\ge 4$ the correlations are essentially always non-local unless the full ground state is separable. 
In Appendix D we also show how the~$N$-spin LHV models perform when restricted to a two-spin subsystem.

\section{Conclusion}

We have introduced an algorithm for the automatic construction of LHV models which reproduce the measurement statistics of quantum many-body states. Importantly, this includes arbitrary continuous measurement settings. We demonstrated the effectiveness of the algorithm by applying it to projective measurements of spin-$1/2$ systems. For two-qubit Werner states as well as noisy three-qubit GHZ and W states we were able for the first time to present estimates for their critical visibilities. Furthermore, we established that two-spin subsystems of the ground states in nearest-neighbor XXZ lattice models are always local. On the other hand, $N$-spin sub-chains in a one dimensional chain appear to be non-local for~$N\ge 4$, in particular around the gapless-to-anti-ferromagnetic phase transition.

We speculate that these properties are not specific to XXZ models. As mentioned before, it is already known that two-spin subsystems in the ground state of translationally invariant Hamiltonians never violate the CHSH inequality. Based on our results one may expect them to be genuinely local as well. On the other hand, we observe that larger subsystems are in general not local, unless the ground state is separable.

While the method is numerical and therefore approximate in nature, it can be made arbitrarily accurate in a systematic way. The local measurement rule becomes more and more general as one increases the maximal degree~$D$ of the spherical harmonics; increasing the hidden-state cloud size~$N_{\text{h}}$ allows us to represent ever more complex hidden-state distributions accurately. Taking both limits together we can represent arbitrary LHV models. 

Our algorithm readily enables the study of locality in all kinds of situations. For example, one could calculate the critical temperature at which thermal states of many-body systems become local. Also, when comparing against analytical methods, we are not restricted to \lq nice\rq\ states such as those with a high degree of symmetry. Hence, it is now straightforward to obtain the critical visibilities for more complicated states or noise channels more elaborate than the fully depolarizing noise studied for Werner states. Finally, in special situations the CHSH inequality is not violated even during non-equilibrium time evolution. With the help of the method developed here, it is now possible to answer the question whether the corresponding states are also local.

\section{Data and Code Availability}

The implementation of the algorithm in Python (we use JAX) along with an example Jupyter notebook demonstrating its application is available on GitHub \cite{github}. There we also provide the figures and corresponding data files.

\bibliography{bibliography}

\onecolumngrid

\appendix

\section{Appendix}

\section{A. Accuracy of the Method}

The presented approach is numerical and thereby approximate to some degree. We evaluate its accuracy and justify how we interpret the results.

\begin{figure}[!ht]
    \centering
    \includegraphics[width=0.99\textwidth]{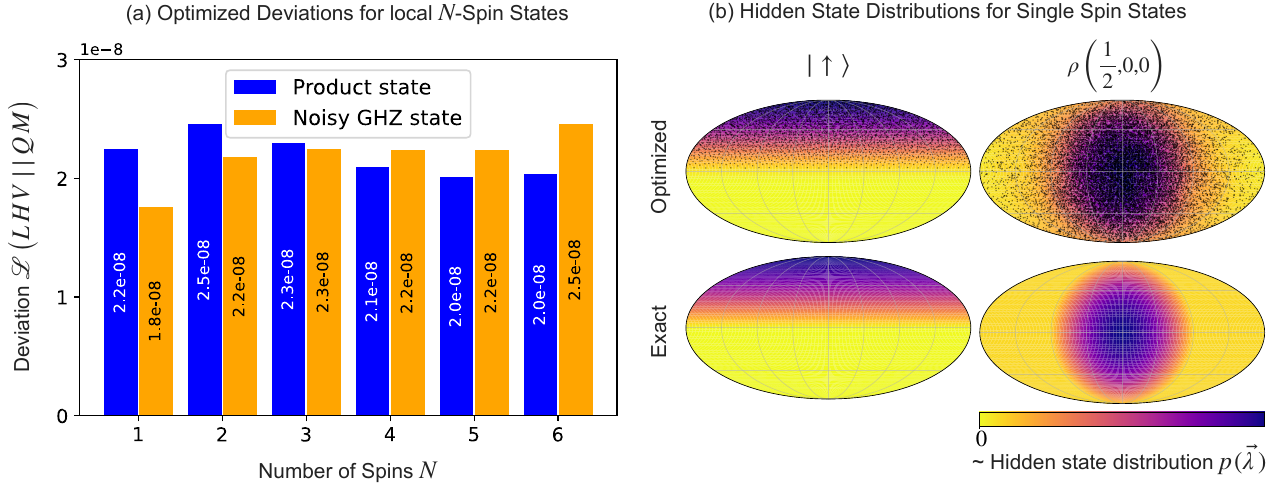}
    \caption{
    (a): Optimized deviation~$\mathcal{L}(LHV||QM)$ for local states (product states and noisy GHZ states) and different numbers of spins~$N$. We consistently obtain~$\mathcal{L} \approx 2 \cdot 10^{-8}$ to~$2.5 \cdot 10^{-8}$.
    (b): Exact (bottom) and optimized (top) hidden-state distributions for a single spin in the state~$\ket{\uparrow}$ (left) and~$\rho(\frac{1}{2}, 0, 0)$ (right) using Bell's measurement rule. We normalize the hidden states to the surface of a sphere and visualize the latter by an area preserving (Mollweide) projection. The black dots are the actual hidden states making up the cloud. The color density is obtained by a Gaussian kernel estimate. For the mixed state we add noise to the gradient descent update steps (\lq diffusion steps\rq) in the first half of the training phase. This ensures a smoother solution. Otherwise a different, more oscillatory (but equally valid) solution is found.}
    \label{fig: baseline and single spin}
\end{figure}

First, we explain the particular value of the optimized loss for local states. Additionally to the examples considered in the main text we optimize LHV models for product states and noisy GHZ states of different numbers of spins. We show the optimized losses in Fig.~\ref{fig: baseline and single spin} (a). As in all examples of local states studied so far we consistently get values around~$\mathcal{L} \approx 2 \cdot 10^{-8}$ to~$3 \cdot 10^{-8}$. We can explain this particular value from the combination of two things: We work with single precision floats (\lq float32\rq) since this is efficient computationally. And we chose the Kullback-Leibler divergence as the loss function. If one only optimizes a single scalar to match a target value using gradient descent with the KL-divergence as loss function and with single precision floats, the loss will still plateau at roughly the same value. Thus we are assured the observed behaviour is not an artifact of the rest of the algorithm, in particular not of the way we represent the hidden-state distribution. Note that the precision of the mantissa in float32 numbers is of the same order (~$2^{-24}\approx 6\cdot 10^{-8}$). Changing the datatype (e.g. to doubles or shorts) or the loss function (e.g. to a square difference loss) will in general change this value.

Next, we consider two situations where the result is known.
Using Bell's measurement rule we have shown that there exist hidden-state distributions reproducing the single spin measurement statistics exactly and we know what these distributions are (see section J below). We find that they are correctly found by our optimization scheme: As a demonstration consider a single spin in either the state~$\ket{\uparrow}$ or the state~$\rho(\frac{1}{2}, 0, 0)$. The optimized hidden-state clouds are shown in Fig.~\ref{fig: baseline and single spin} (b). Visually, the optimized distributions coincide well with the exact results and the optimized deviation is again roughly~$2\cdot 10^{-8}$ to~$3\cdot 10^{-8}$.

\begin{figure}[!ht]
    \centering
    \includegraphics[width=0.99\textwidth]{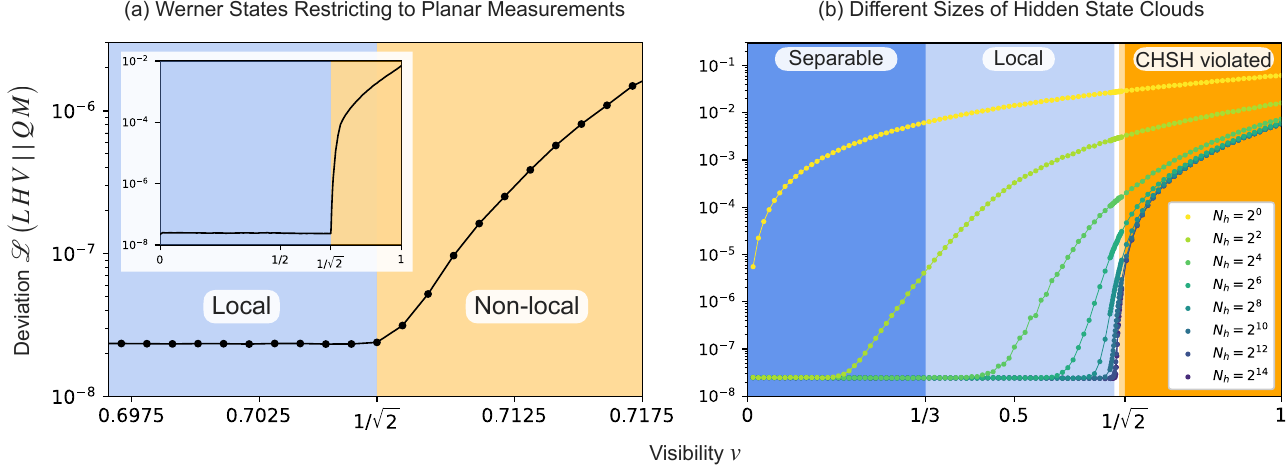}
    \caption{
    (a): Optimized deviation~$\mathcal{L}(LHV||QM)$ for Werner states measured in a plane. The critical visibility in this case is~$v_C= 1/\sqrt{2}$. The deviation stays small and constant up to the critical visibility and increases thereafter. 
    (b): Optimized deviation~$\mathcal{L}(LHV||QM)$ for Werner states (now again under arbitrary projective measurements). As the size~$N_{\text{h}}$ of the hidden-state cloud increases, the solution converges.}
    \label{fig: Werner N_hidden and planar}
\end{figure}

If one restricts to measurements in a plane, the critical visibility for Werner states is known exactly \cite{Toner07}
\begin{align}
    v_C^{\text{(2D)}} = \frac{1}{\sqrt{2}}.
\end{align}
To analyze this scenario, we optimize LHV models for the Werner states while only sampling measurement directions in the~$xy$-plane. The result is shown in Fig.~\ref{fig: Werner N_hidden and planar} (a). We observe the transition to increasing deviations at the correct critical visibility. This gives us confidence that our previous estimate for the critical visibility in case of all projective measurements is accurate as well.

Finally we analyze the dependence of the solution on the size of the hidden-state cloud~$N_{\text{h}}$ (Fig.~\ref{fig: Werner N_hidden and planar} (b)). Larger clouds can represent more complex distributions and approximate continuous distributions more accurately. We optimize LHV models for the Werner states (now again for all projective measurements) using different~$N_{\text{h}}$. As expected, we see that the optimized deviations converge to some lower bound as we increase~$N_{\text{h}}$. Already for small clouds we reach~$\mathcal{L}\approx 2.5\cdot 10^{-8}$ for small visibilities. For larger clouds the region of small loss grows. This corresponds to a convergence of the estimate for~$v_C$ to some point in the previously unknown region. Increasing the maximal degree in~$D$ in the spherical-harmonics expansion leads to similar behaviour.

\subsection{B. Glassy Optimization Landscape and Hyper Parameters}

\begin{figure}[!ht]
    \centering
    \includegraphics[width=0.5\textwidth]{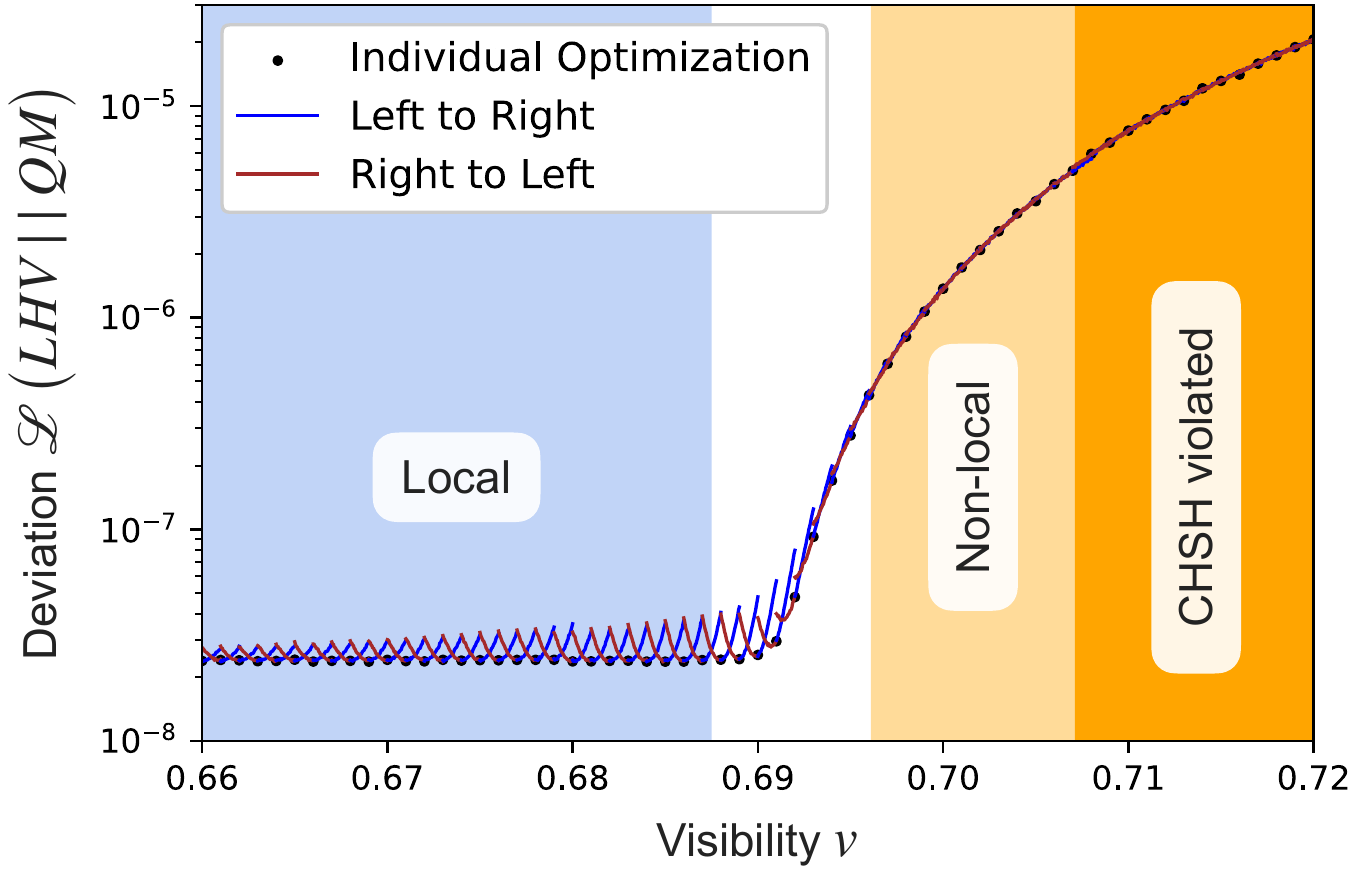}
    \caption{Optimized deviation~$\mathcal{L}(LHV||QM)$ for Werner states (black dots) using a hidden-state cloud of size~$N_{\text{h}}=2^{14}$. For each pair of visibilities we \lq interpolate\rq\ in between by always using the result for the closest visibility as initialization (blue and red lines). Far from the critical visibility this yields the same results as with independent initialization. There the solution seems to depend on the visibility in a continuous way. Close to the critical visibility however, the results obtained in this way are significantly worse compared to independent initialization. This indicates a non-continuous dependence of the optimal hidden-state cloud on the visibility.}
    \label{fig: Werner interpolate}
\end{figure}

We already mentioned in the main text that convergence of the hidden-state distribution takes longer close to the critical visibility of Werner states. We observe similar behaviour for different kinds of states always at the boundary between local and non-local states. It seems to be easy for our class of LHV models to represent very local states, such as separable ones. Conversely, for highly entangled, very non-local states there is not much to optimize (the approximation remains bad). Hence in these cases convergence is fast. In between, close to the boundary, the optimization landscape seems to be very complicated or \lq glassy\rq\ in the sense that the optimal hidden-state distribution depends on, e.g., the visibility in a non-continuous way. This can be seen by optimizing the hidden-state cloud for some visibility~$v$ and using the result as initialization for~$v' = v\pm\Delta v$. We observe that close to~$v_C$ the results obtained in this manner for~$v'$ are significantly worse compared to optimizing with unbiased initialization (i.e. fresh initialization for each value of the visibility). In detail, we individually optimize LHV models for a range of Werner states. In a second step we \lq interpolate\rq\ in between each pair of neighboring visibilities by always using the result for the closest visibility as initialization. This is shown in Fig.~\ref{fig: Werner interpolate}. We interpret these observations as meaning that the loss function has many local minima whose relative depths depend on the visibility, thereby constantly changing the global minimum in a non-continuous way.

The hyper-parameters which change the run-time of the algorithm are (together with the typical values we used)
\begin{itemize}
    \item Number of gradient descent steps~$N_{\text{steps}} = 10^4\ldots 10^7$
    \item Number of spins~$N=1\ldots 6$
    \item hidden-state cloud size~$N_{\text{h}}=2^{10}\ldots 2^{16}$
    \item Maximal degree~$D=1\ldots 5$ in the spherical-harmonics expansion of the measurement rule
    \item Dimension of the hidden states (per spin)~$d = \frac{1}{2}(D+1)(D+2)$
    \item The batch size~$N_{\text{batch}}=2^8\ldots 2^{10}$, that is the number of combinations of measurement directions sampled per gradient descent step.
\end{itemize}
For completeness we mention the learning rate~$\text{lr}$ with typical values~$\text{lr}/N_{\text{h}} = 5\cdot 10^{-6}\ldots 5\cdot 10^{-5}$. 
To give a specific example for the run-time, with hyper parameters~$N_{\text{steps}}=10^4,\ N=2,\ N_{\text{h}}=2^{14},\ D=5,\ d=21,\ N_{\text{batch}} = 2^{10}$ sampling all projective measurements for a Werner state, the run-time on a single NVIDIA Quadro RTX 6000 GPU is about~$16$ seconds ($18$ seconds when running the jitted function for the first time).  Optimizing multiple states in parallel allows for an additional speed up. Generating the data for Fig.~\ref{fig: Werner} takes a few days with most of the time spent on the visibilities close to~$v_C$.

\subsection{C. GHZ and W States}

\begin{figure}[!ht]
    \centering
    \includegraphics[width=0.99\textwidth]{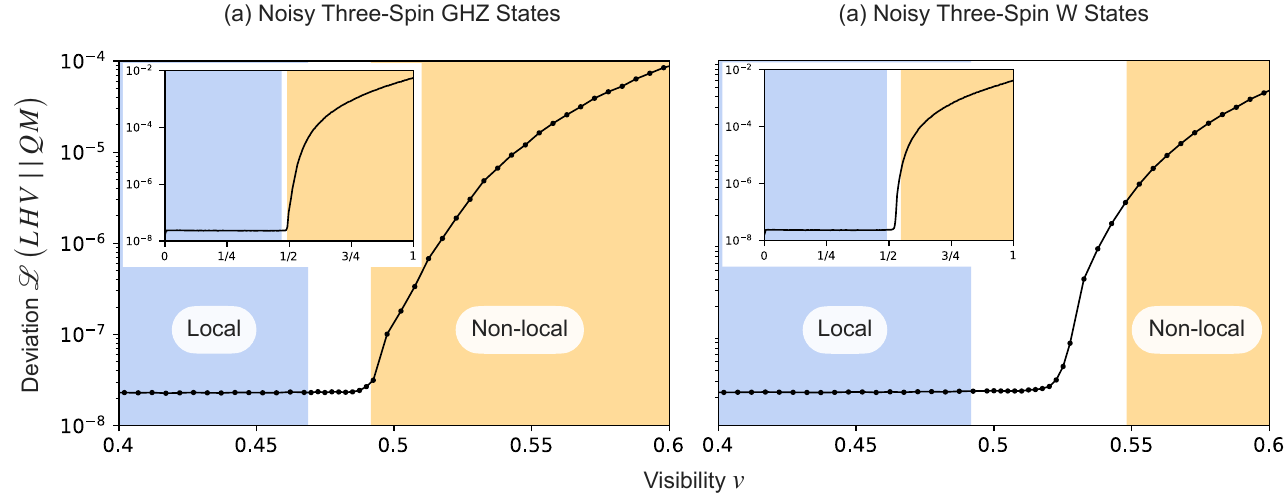}
    \caption{Optimized deviation~$\mathcal{L}(LHV||QM)$ for noisy GHZ (a) and W (b) states. Compared to Werner states the regions where it is unknown whether they are local or not are even larger. In both cases we again obtain estimates for the critical visibilities. For the GHZ states~$v_C\approx 0.485$ which lies towards the right boundary of the previously unknown region. For the W states~$v_C\approx 0.513$ which lies towards the center of the previously unknown region.}
    \label{fig: GHZ and W}
\end{figure}

Werner states are noisy Bell singlets. Analogously, we can consider noisy three-qubit GHZ and W states
\begin{align}
    \ket{GHZ} = \frac{1}{\sqrt{2}}\left( \ket{\uparrow\uparrow\uparrow} + \ket{\downarrow\downarrow\downarrow}\right),\qquad \ket{W} = \frac{1}{\sqrt{3}}\left(\ket{\downarrow\uparrow\uparrow} + \ket{\uparrow\downarrow\uparrow} + \ket{\uparrow\uparrow\downarrow}\right)
\end{align}
Similar to the two-qubit Werner states, only bounds for their critical visibilities are known \cite{Designolle23}. We optimize LHV models for both of them and again find numerical estimates for their critical visibilities (see Fig.~\ref{fig: GHZ and W})
\begin{align}
    v_C^{GHZ} \approx 0.485,\qquad v_C^{W} \approx 0.518.
\end{align}

\subsection{D. More on XXZ Groundstates}

\begin{figure}[!ht]
    \centering
    \includegraphics[width=0.99\textwidth]{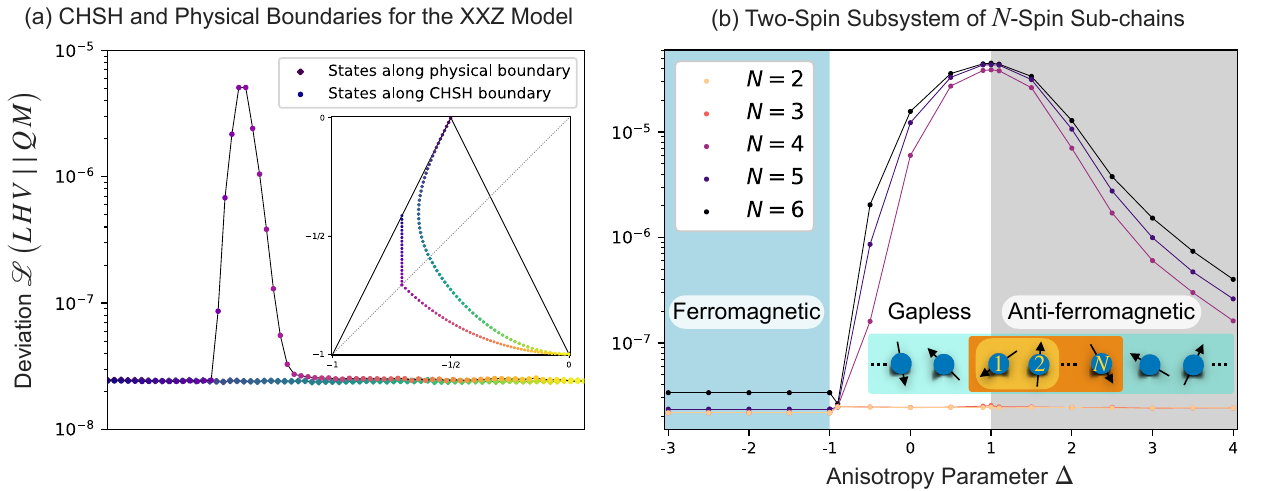}
    \caption{
    (a) Optimized deviation~$\mathcal{L}(LHV||QM)$ for the two-spin states with XXZ symmetries along the CHSH boundary (circles) and the physical boundary (diamonds). Along the CHSH boundary the deviation is small and constant everywhere except close the the Werner states. Along the physical boundary the deviation is small and constant everywhere. This implies that all two spin subsystems of the ground states of XXZ models on lattices with arbitrary geometries are local.
    (b) Deviation~$\mathcal{L}(LHV||QM)$ for two-spin subsystems of~$N$-spin sub-chains in the ground state of a one-dimensional XXZ model. The deviation was optimized on the~$N$-spin sub-chains. If the deviation was small for the latter it is also small on its two-spin subsystem. If the deviation was larger on the~$N$-spin chain we also get a larger deviation on the two-spin subsystem, even though it by itself is always local.
    }
    \label{fig: XXZ extra}
\end{figure}

In the main text we claimed that for the two-qubit states respecting the symmetries of the XXZ Hamiltonian the CHSH inequality is almost tight, except close to the Werner states. To see this more clearly, we optimize LHV models along the boundary, that is along the dashed and the red line in Fig.~\ref{fig: XXZ 2D}. This is shown in Fig.~\ref{fig: XXZ extra} (a). We can clearly see that along the physical boundary the deviation stays small throughout. Along the CHSH boundary the deviation is small except close to the Werner states where we find a local maximum.

So far we spoke of \lq all projective measurements\rq\ for the spin-$1/2$ systems. In practice, we only considered the non-trivial ones, that is excluding the option of not measuring a spin. However, if all non-trivial measurements are correctly captured by the LHV model we automatically get the correct result for the trivial measurements as well. To check how well this works in practice we take the optimized models for the~$N$-spin sub-chains of the XXZ ground state in one dimension (Fig.~\ref{fig: XXZ 1D}) and test them on just two spins (Fig.~\ref{fig: XXZ extra} (b)). We observe that our expectation is essentially correct: In case the~$N$-spin subsystem is local and we found an accurate LHV model, it also performs well on the two spin subsystem. In case the~$N$-spin subsystem is non-local and the optimized LHV model is bad it also performs poorly on the subsystem.

\subsection{E. Remark on Observables}

Suppose a model of the form in eq.~(\ref{eq: LHV correlations physical}) describes the measurement statistics of a given $N$-particle quantum state~$\rho$ exactly for all projective measurements. Consider any observable~$A$ for the~$N$ particles. It corresponds to a hermitian operator which can be decomposed as a linear combination of products of single particle projection operators
\begin{align}
    A = \sum_k c_k P^{(k)}_1\tp \cdots \tp P^{(k)}_N.
\end{align}
Therefore, the quantum mechanical expectation value can be calculated using the LHV model
\begin{align}
    \notag
    \langle A \rangle &
    = \sum_k c_k \Tr(P^{(k)}_1\tp \cdots \tp P^{(k)}_N \cdot \rho)
    = \sum_k c_k P^{LHV}(P^{(k)}_1,\ldots, P^{(k)}_N \, | \, \rho).
\end{align}
The decomposition of~$A$ is not unique. There are infinitely many choices for such representations. As long as all measurement statistics are exactly reproduced by the LHV model, this is not a problem. However, in practice this will only hold approximately when we represent non-local states using an LHV model (also, strictly speaking, even in the local region due to small remaining approximation errors). In this situation the LHV statistics generically do not represent a quantum state. Then, the expectation value predicted by the LHV model depends on the decomposition of the observable. 
The simplest way to see that there are LHV models which are incompatible with quantum mechanics is for a single qubit:
For any qubit state and two measurement directions~$\hat n_1,\hat n_2\in S^2$ the quantum mechanical probabilities for obtaining \lq up\rq\ are not independent in the sense that
    \begin{align}
        \label{eq: PQM restriction}
        P^{QM}(\hat n_1) + P^{QM}(\hat n_2) \le 1 + \abs{\cos(\theta/2)},\qquad \hat n_1\cdot \hat n_2 = \cos(\theta).
    \end{align}

\begin{proof}

    Let the state be~$\rho = \rho(\vec r)$, then
    \begin{align}
        P^{QM}(\hat n_1) + P^{QM}(\hat n_2) &= \frac{1}{2}\left(1 + \hat n_1\cdot \vec r\right) + \frac{1}{2}\left(1 + \hat n_2\cdot \vec r\right) = 1 + \frac{1}{2}(\hat n_1 + \hat n_2)\cdot \vec r \\
        &\le 1 + \frac{1}{2}\norm{\hat n_1 + \hat n_2}\norm{\vec r} \le 1 + \frac{1}{2}\norm{\hat n_1 + \hat n_2} \\
        &= 1 + \frac{1}{2}\sqrt{\norm{\hat n_1}^2 + \norm{\hat n_2}^2 + 2 \hat n_1\cdot \hat n_2} 
        = 1 + \frac{1}{\sqrt{2}} \sqrt{1 + \hat n_1\cdot \hat n_2} \\
        &= 1 + \abs{\cos(\theta/2)}.
    \end{align}
\end{proof}

For distinct measurement directions the sum of probabilities is strictly less than two. However, our class of LHV models contains hidden-state distributions that for any given pair of measurement directions with relative angle less than~$\pi$ allow for this sum to be equal to two. This means there are LHV models whose correlations do not correspond to any state. One could try to constrain the hidden-state distributions, however it is not clear how to do this.

\subsection{F. Different Definitions of Bell-Locality}

The two presented ways to construct Bell-local correlations are equivalent in the sense that correlations can be written in the form of eq.~(\ref{eq: LHV correlations standard}) if and only if they can be written in the form of eq.~(\ref{eq: LHV correlations physical}).

\begin{proof}
    Suppose~$P(a|x)$ can be written in the form of eq.~(\ref{eq: LHV correlations standard}). Define the new hidden-state space
    \begin{align}
        \Lambda' = \Lambda \times \{1,\ldots, N\} \ni \lambda' = (\lambda, n).
    \end{align}
    Set the hidden-state distribution to be
    \begin{align}
        p'(\lambda_1', \ldots, \lambda_N') = \delta(\lambda_1-\lambda_2)\cdots \delta(\lambda_1-\lambda_N) \delta_{n_1, 1}\cdots \delta_{n_N, N} p(\lambda_1)
    \end{align}
    and the local measurement rule to
    \begin{align}
        q'(a|x, (\lambda, n)) = q_n(a|x, \lambda).
    \end{align}
    Then
    \begin{align}
        P'(a|x) &= \sum_{n_1=1}^N\int_{\Lambda}\dup \lambda_1 \cdots \sum_{n_N=1}^N\int_{\Lambda}\dup \lambda_N\, \delta(\lambda_1-\lambda_2)\cdots \delta(\lambda_1-\lambda_N) \delta_{n_1, 1}\cdots \delta_{n_N, N} p(\lambda_1) \\
        &\hspace{150pt}\times q_{n_1}(a_1|x_1, \lambda_1)\cdots q_{n_N}(a_N|x_N, \lambda_N) \\
        &= \int \dup \lambda_1\, p(\lambda_1) q_1(a_1|x_1, \lambda_1)\cdots q_N(a_N|x_N, \lambda_1) 
        = P(a|x).
    \end{align}
    Now, suppose~$P(a|x)$ can be written in the form of eq.~(\ref{eq: LHV correlations physical}). Define the new hidden-state space
    \begin{align}
        \Lambda' = \Lambda^N \ni \lambda' = (\lambda_1,\ldots, \lambda_N)
    \end{align}
    with hidden-state distribution
    \begin{align}
        p'(\lambda') = p(\lambda_1,\ldots, \lambda_N)
    \end{align}
    and local measurement rules
    \begin{align}
        q_j(a|x, \lambda') = q(a|x, \lambda_j).
    \end{align}
    Then
    \begin{align}
        P'(a|x) &= \int_{\Lambda^N}\dup \lambda'\, p'(\lambda') q_1(a_1|x_1, \lambda')\cdots q_N(a_N|x_N, \lambda')\\
        &= \int_{\Lambda}\dup \lambda_1 \cdots \int_\Lambda \dup\lambda_N\, p(\lambda_1,\ldots, \lambda_N) q(a_1|x_1, \lambda_1)\cdots q(a_N|x_N, \lambda_N) = P(a|x).
    \end{align}
\end{proof}

\subsection{G. Non-Contextual Measurement Rules}

Besides the hidden-state vector~$\lambda$ the local measurement rule depends on the measurement input~$x$ and the measurement outcome~$a$. A measurement corresponds to a POVM (for simplicity we stick to discrete ones here)
\begin{align}
    x = \{M_k\}_{k=1}^{K},\quad 0\le M_k \le \1,\quad \sum_{k=1}^K M_k = \1,\quad K\in\mathbb{N}\cup \{\infty\}.
\end{align}
And the possible outputs are the POVM elements:~$a=M_k\in x$.
In this sense the local measurement rule is contextual: The probability for a specific outcome can depend on the possible alternative outcomes defined by the measurement setup. However, without loss of generality we can assume non-contextual measurement rules. To see this suppose an~$N$-particle state~$\rho$ admits an LHV model
\begin{align}
    P^{QM}(a|x) = \int_{\Lambda^N}\dup \lambda p(\lambda) \prod_{j=1}^N q(a_j|x_j, \lambda_j) \quad \forall x=(x_1,\ldots,x_N),\ \forall a=(a_1,\ldots,a_N),a_j\in x_j.
\end{align}
Since the quantum mechanical measurement probabilities are non-contextual
\begin{align}
    P^{QM}(a|x) = \Tr(a_1\tp\cdots\tp a_N\cdot \rho) = P^{QM}(a)
\end{align}
we can simply choose an arbitrary valid context. For instance
\begin{align}
    \overline{q}(a , \lambda) \equiv q(a | \{a, \1-a\}, \lambda)\ \Rightarrow\ P^{QM}(a) = \int_{\Lambda^N}\dup \lambda p(\lambda) \prod_{j=1}^N \overline{q}(a_j , \lambda_j) \quad \forall a.
\end{align}
That is,~$\rho$ admits a non-contextual LHV model and it is sufficient to look for such models to determine whether a given state is local or not. If one is interested in a subset of all POVM measurements the same argument applies as long as one can fix a context for all possible measurement outcomes. For spin-$1/2$ systems this shows that we did not loose anything by working (only) with~$q(\hat n,\lambda) = q(\uparrow|\hat n,\lambda)$.

With the choice above (which is in particular valid for projective measurements) we have the normalization 
\begin{align}
    \overline{q}(a,\lambda) + \overline{q}(1-a,\lambda) = q(a,\{a, \1-a\},\lambda) + q(1-a,\{a, \1-a\},\lambda) = 1
\end{align}
which translates to~eq.~(\ref{eq: spin measurement rule constraint}) under the change of notation $a=P_{\hat n}\mapsto \hat n$, $\1-P_{\hat n} = P_{-\hat n} \mapsto -\hat n$ ($P_{\hat n}$ is the projection onto the \lq up\rq\ eigenstate of~$\hat n\cdot \vec\sigma$).  We note that this is slightly more than \lq just normalization\rq: It implicitly assumes that measuring \lq up\rq\ along direction~$\hat n$ is the same as measuring down in direction~$-\hat n$. It is not obvious why this would hold for individual instances of the hidden variable. However, the argument above shows that we can assume so.

\subsection{H. Efficient Loss Function}

Our initially proposed loss function for spin-$1/2$ systems would be the average of the KL-divergence between~$P^{QM}$ and~$P^{LHV}$ over all possible combinations of measurement directions~$x=(\hat n_1,\ldots,\hat n_N)$. That is, the average of
\begin{align}
    \text{D}_{KL}\hspace{-2pt}\left(P^{QM}(\cdot|x) || P^{LHV}(\cdot|x)\right)
    &= \sum_{s_1,\ldots,s_N\in\{\pm 1\}} P^{QM}(s_1\hat n_1,\ldots , s_N \hat n_N) \ln(\frac{P^{QM}(s_1\hat n_1,\ldots , s_N \hat n_N)}{P^{LHV}(s_1\hat n_1,\ldots , s_N \hat n_N)})
\end{align}
This is an exponentially large sum in the number of spins. A more efficient alternative is to only consider the outcomes \lq all spins up\rq\ and \lq not all spins up\rq. That is, we consider the KL-divergence between the simplified distributions~$\{P^{QM}(x), 1-P^{QM}(x)\}$ and~$\{P^{LHV}(x), 1-P^{LHV}(x)\}$ (we write $P(x) \equiv P(\uparrow\cdots \uparrow | x)$)
\begin{align}
    \mathcal{L}({LHV}\, ||\,{QM}) = \left\langle P^{QM}(x)\ln(\frac{P^{QM}(x)}{P^{LHV}(x)}) + \left(1-P^{QM}(x)\right)\ln(\frac{1-P^{QM}(x)}{1-P^{LHV}(x)})\right\rangle_x.
\end{align}
This is enough because we sample all possible measurement directions. A vanishing loss means~$P^{QM}(x) = P^{LHV}(x)$ for all~$x$, i.e. the LHV model reproduces quantum mechanics perfectly. For non-local states this change of the loss function can lead to different compromises for the optimal hidden-state distribution as mentioned before.

\subsection{I. POVM Measurements for Spin-1/2 Particles}
\label{App-sec: povm measurements for spin-1/2 particles}

In this appendix, we work out how to deal with general POVM measurements. We have not implemented this so far in our numerics, though.
The most general single qubit POVM element is a positive two by two matrix bounded by the identity~$0 \le M \le \1$. A general positive two by two matrix can be obtained by scaling a single qubit state ($\vec r$ is a Bloch vector)
\begin{align}
    M = m\rho(\vec r) = \frac{m}{2} \left(\1 + \vec r\cdot \vec \sigma\right),\qquad m \ge 0,\ \norm{\vec r}\le 1.
\end{align}
The condition~$M\le \1$ can then be implemented by requiring
\begin{align}
    m\le \frac{2}{1 + \norm{\vec r}}.
\end{align}
This can be rewritten as ($m_0 = m/2, \vec m = m \vec r$)
\begin{align}
    M = M(m_\mu) = \sum_{\mu} m_\mu \sigma^\mu,\qquad \norm{\vec m} \le \min(m_0, 1-m_0),
\end{align}
where~$\sigma^\mu = (\1,\vec \sigma)$ and implicitly~$0 \le m_0 \le 1$. The projection operators correspond to the special case~$m_0 = 1/2$. We have
\begin{align}
    \1 - M(m_0, \vec m) = M(1-m_0, -\vec m).
\end{align}
For implementations it will be convenient to rewrite this expression by setting $m_0^{\rm old} = m_0^{\rm new} + 1/2$  such that
\begin{align}
    M(m_\mu) = \left(\frac{1}{2} + m_0\right)\1 + \vec m\cdot \vec\sigma,\qquad \abs{m_0} \le \frac{1}{2},\ \norm{\vec m} \le \frac{1}{2} - \abs{m_0}.
\end{align}
In this case
\begin{align}
    \1 - M(m_\mu) = M(-m_\mu),
\end{align}
which means the local measurement rule can again be written as
\begin{align}
    q(m_\mu , \lambda) = \sigma(f_{\lambda}(m_\mu))
\end{align}
with odd functions~$f_{\lambda}$. The mean over POVM elements in the loss function can just be taken with respect to a uniform distribution over the allowed (compact) range of~$m_\mu$. The functions~$f_\lambda$ can be expanded into odd polynomials where the coefficients again become the components of the hidden-state vector.

\subsection{J. Exact LHV Models for Separable States}

If a given measurement rule~$q$ allows to represent all single particle states ($M$ is a POVM element)
    \begin{align}
        \forall \rho\, \exists\, p(\lambda|\rho):\ \Tr(M\rho) = \int_\Lambda\dup\lambda\, q(M|\lambda) p(\lambda|\rho),
    \end{align}
    then all non-entangled $N$-particle states can be represented exactly.
    
\begin{proof}
    Any separable $N$~particle state is a convex linear combination of product states
    \begin{align}
        \rho = \sum_{k}t_k \rho^{(k)}_1\tp \cdots \tp \rho^{(k)}_N,\qquad \sum_k t_k = 1.
    \end{align}
    By assumption we have exact hidden-state distributions~$p(\lambda | \rho^{(k)}_j)$ for the individual parts~$\rho^{(k)}_j$. We claim that
    \begin{align}
        p(\lambda_1,\ldots,\lambda_N|\rho) = \sum_k t_k p(\lambda_1|\rho^{(k)}_1)\cdots p(\lambda_N|\rho^{(k)}_N)
    \end{align}
    is the correct hidden-state distribution. Indeed this reproduces the correct measurement statistics
    \begin{align}
        &P^{LHV}(M_1,\ldots,M_N |\rho) \\
        &= \int_\Lambda \dup \lambda_1\cdots \int_\Lambda\dup\lambda_N\,p(\lambda_1,\ldots,\lambda_N|\rho) q(M_1,\lambda_1)\cdots q(M_N,\lambda_N)\\
        &= \sum_k t_k \int_\Lambda\dup\lambda_1\, p(\lambda_1|\rho^{(k)}_1) q(M_1,\lambda_1)\cdots \int_\Lambda\dup\lambda_N\, p(\lambda_N|\rho^{(k)}_N) q(M_N,\lambda_N)\\
        &= \sum_k t_k P^{LHV}(M_1| \rho^{(k)}_1)\cdots P^{LHV}(M_N| \rho^{(k)}_N) \\
        &= \sum_k t_k \Tr(M_1\rho^{(k)}_1)\cdots \Tr(M_N\rho^{(k)}_N)\\
        &= \Tr(\sum_k t_k M_1\tp\cdots\tp M_N \cdot \rho^{(k)}_1\tp \cdots \tp \rho^{(k)}_N) \\
        &= \Tr(M_1\tp\cdots \tp M_N \cdot \rho) = P^{QM}(M_1,\ldots,M_N|\rho).
    \end{align}
\end{proof}

And with Bells measurement rule -- and thereby with the spherical harmonic rule of any degree greater or equal to one -- at least all separable $N$~spin-$1/2$ states can be represented exactly with respect to projective measurements. The hidden-state distribution for a single spin in the state~$\rho = \rho(\vec r)$ ($\vec r$ is the Bloch vector) is given by
    \begin{align}
        p(\hat \lambda | \vec r)  &= \frac{1}{\pi}\left( \vec r \cdot \hat \lambda\, H(\vec r \cdot \hat \lambda) + \frac{1-\abs{\vec r}}{4} \right).
    \end{align}
\begin{proof}

    We have
    \begin{align}
        P^{LHV}_\uparrow(\hat n) = \int_{S^2}\dup A(\hat\lambda)\, p(\hat \lambda) H(\hat n\cdot \hat \lambda).
    \end{align}
Let us simplify this for a general hidden-state distribution~$p(\hat \lambda) = p(\tilde \phi, \tilde\theta)$
\begin{align}
    \notag
    P(\phi, \theta) &\equiv P_\uparrow^{LHV}(\hat n(\phi, \theta)) = \int_{S^2}\dup A(\hat\lambda)\, p(\hat \lambda) H(\hat n(\phi, \theta)\cdot \hat \lambda), \\ \notag
    P(\phi, 0) &= \int_{0}^{2\pi}\dup \tilde\phi \int_{0}^1 \dup \cos\hspace{0pt}(\tilde \theta)\, p(\tilde\phi, \tilde\theta),\\ 
    \partial_\theta P(\phi, \theta) &= \int_{S^2}\dup A(\hat\lambda)\, p(\hat \lambda) \partial_\theta(\hat n \cdot \hat \lambda) \delta(\hat n \cdot \hat \lambda).
\end{align}
Introduce
\begin{align}
    \hat n_x = \begin{bmatrix} \cos(\theta)\cos(\phi) \\ \cos(\theta)\sin(\phi) \\ -\sin(\theta) \end{bmatrix} = \partial_\theta \hat n, \qquad \hat n_y = \begin{bmatrix} -\sin(\phi) \\ \cos(\phi) \\ 0 \end{bmatrix}.
\end{align}
Together with~$\hat n_z = \hat n$ they form a right-handed orthonormal basis. The delta function~$\delta(\hat n\cdot \hat \lambda)$ means the integration becomes restricted to the circle (centered at the origin) that is orthogonal to~$\hat n$, that is the circle around the origin in the~$\hat n_x,\,\hat n_y$ plane. We can parameterize this circle as
\begin{align}
    \notag
    \hat \lambda(\tau) &= \cos(\tau)\hat n_x + \sin(\tau)\hat n_y,\quad \tau \in [0,\, 2\pi], \\
    \norm*{\dot{\hat {\lambda}}(\tau)}^2 &= \norm{-\sin(\tau)\hat n_x + \cos(\tau)\hat n_y}^2 = 1.
\end{align}
Then
\begin{align}
    \notag
    \partial_\theta P(\phi, \theta) &= \int_{0}^{2\pi}\dup \tau\, p(\hat \lambda(\tau)) \left(\hat n_x\cdot \hat \lambda(\tau)\right) \norm*{\dot{\hat {\lambda}}(\tau)} 
    = \int_0^{2\pi}\dup \tau \, \cos(\tau) p(\hat\lambda(\tau)) .
\end{align}
Plugging in the specific trajectory we obtain
\begin{align}
    \partial_\theta P(\phi, \theta) &= \int_0^{2\pi}\dup \tau\, \cos(\tau)\, p\hspace{-3pt}\left( \begin{bmatrix}
        \cos(\tau)\cos(\theta)\cos(\phi) - \sin(\tau)\sin(\phi) \\
        \cos(\tau)\cos(\theta)\sin(\phi) + \sin(\tau)\cos(\phi) \\
        -\cos(\tau)\sin(\theta)
    \end{bmatrix} \right)
\end{align}
Consider a general single spin state
\begin{align}
    \rho = \rho(\vec r) = \frac{1}{2}\left(\1 + \vec r\cdot \vec \sigma\right).
\end{align}
This yields 
\begin{align}
    P^{QM}_\uparrow(\hat n) = \frac{1}{2}\left(1 + \vec r \cdot \hat n\right).
\end{align}
We have (and this uniquely determines~$P^{QM}_\uparrow(\phi, \theta)$)
\begin{align}
    \notag
    P_\uparrow^{QM}(\phi, \theta = 0) &= \frac{1}{2}\left(1 + r_z\right), \qquad
    \partial_\theta P_\uparrow^{QM}(\phi, \theta) = \frac{1}{2}\vec r \cdot \hat n_x.
\end{align}
First consider the case~$\rho = \ketbra{0}$, that is~$\vec r = \hat e_z = (0, 0, 1)$. Then
\begin{align}
    P^{QM}_\uparrow(\phi, \theta = 0) = 1,\qquad \partial_\theta P_\uparrow^{QM}(\phi, \theta) = -\frac{1}{2}\sin(\theta). 
\end{align}
We claim that the correct hidden-state distribution which reproduces this is (with~$\hat \lambda = \hat \lambda(\tilde \phi, \tilde \theta)$)
\begin{align}
    p(\hat \lambda | \vec r = \hat e_z) = \frac{\hat e_z \cdot \hat \lambda}{\pi} H(\hat e_z \cdot \hat \lambda) = \frac{\cos\hspace{0pt}(\tilde\theta)}{\pi}H(\cos\hspace{0pt}(\tilde\theta)) = \frac{\lambda_z}{\pi}H(\lambda_z).
\end{align}
First of all this is a valid probability distribution over the sphere
\begin{align}
    \int_{S^2}\dup A(\hat \lambda)\, p(\hat \lambda| \hat e_z) &= \int_0^{2\pi}\dup \tilde\phi \int_{-1}^1 \dup (\cos\hspace{0pt}(\tilde \theta))  \frac{\cos\hspace{0pt}(\tilde\theta)}{\pi}H(\cos\hspace{0pt}(\tilde\theta))
    =\frac{2\pi}{\pi}\int_0^1 \dup u\, u = 1.    
\end{align}
Next, the value at~$\theta = 0$ is correct (this is the same integral as the normalization above)
\begin{align}
    P(\phi, 0) &= \int_{0}^{2\pi}\dup \tilde\phi \int_{0}^1 \dup (\cos\hspace{0pt}(\tilde \theta))\, p(\tilde\phi, \tilde\theta | \hat e_z) = 1.
\end{align}
And lastly, the derivative is correct
\begin{align}
    \notag
    \partial_\theta P(\phi, \theta) &=
    \int_0^{2\pi}\dup \tau \, \cos(\tau) p(\hat\lambda(\tau)|\hat e_z) \\ \notag
    &= \int_0^{2\pi}\dup \tau\, \cos(\tau) \left(-\frac{1}{\pi}\cos(\tau)\sin(\theta) H(-\cos(\tau)\sin(\theta))\right) \\ \notag
    &= -\frac{1}{\pi}\sin(\theta) \int_0^{2\pi} \dup \tau\, \cos^2(\tau)H(-\cos(\tau)) \\
    &= -\frac{1}{\pi}\sin(\theta) \int_{\pi/2}^{3\pi/2}\dup \tau\, \cos^2(\tau) = -\frac{1}{2}\sin(\theta)
\end{align}
because~$\sin(\theta)\ge 0$ as~$\theta\in [0,\, \pi]$.

Now assume~$\hat r$ is any normalized vector corresponding to an arbitrary pure state. It corresponds to the hidden-state distribution~$p(\hat \lambda |\hat r)$ such that
\begin{align}
    P^{QM}_\uparrow(\hat n |\hat r) = P^{LHV}_\uparrow(\hat n | \hat r) = \int_{S^2}\dup A(\hat \lambda)\, p(\hat\lambda |\hat r) H(\hat n\cdot \hat \lambda).
\end{align}
We know that for any rotation matrix~$R$ it holds that
\begin{align}
    P^{QM}_\uparrow(\hat n | R\hat r) = \frac{1}{2}(1+(R\hat r)\cdot \hat n) = \frac{1}{2}(1+\hat r\cdot (R^T\hat n)) = P^{QM}_\uparrow(R^T\hat n | \hat r).
\end{align}
Therefore the hidden-state distributions for Bloch vectors on the surface of the Bloch sphere must satisfy
\begin{align}
    \notag
    &\int_{S^2}\dup A(\hat \lambda)\, p(\hat\lambda |R\hat r) H(\hat n\cdot \hat \lambda) 
    \stackrel{!}{=} \int_{S^2}\dup A(\hat \lambda)\, p(\hat\lambda |\hat r) H((R^T\hat n)\cdot \hat \lambda) \\
    &= \int_{S^2}\dup A(R\hat \lambda)\, p(R^T(R\hat\lambda) |\hat r) H(\hat n\cdot (R\hat \lambda)) = \int_{S^2}\dup A(\hat \lambda)\, p(R^T\hat\lambda |\hat r) H(\hat n\cdot \hat \lambda).
\end{align}
This means if~$p(\hat \lambda |\hat r)$ is a valid solution for the Bloch vector~$\hat r$, then~$p(R^T\hat \lambda |\hat r)$ is always a valid solution for the Bloch vector~$R\hat r$ (it is correctly normalized), i.e.
\begin{align}
    p(R^T\hat \lambda | \hat r) = p(\hat \lambda | R\hat r) \quad \forall R, \hat r, \hat \lambda.
\end{align}
Since we know the solution for one unit Bloch vector~($\hat e_z$) we can use this property to obtain the solution for all unit Bloch vectors: Let~$R$ be the rotation matrix mapping~$\hat e_z$ to~$\hat r$, then
\begin{align}
    \notag
    p(\hat \lambda | \vec r) &= p(\hat \lambda |R \hat e_z) = p(R^T\hat \lambda | \hat e_z) =     
    \frac{\hat e_z \cdot (R^T\hat \lambda)}{\pi} H(\hat e_z \cdot (R^T\hat \lambda)) \\
    &= \frac{(R\hat e_z)\cdot \hat \lambda}{\pi} H((R\hat e_z)\cdot \hat \lambda) 
    = \frac{\hat r\cdot \hat \lambda}{\pi} H(\hat r\cdot \hat \lambda).
\end{align}
Finally, we extend the result to non-unit Bloch vectors. Let~$\vec r$ be any Bloch vector and~$\alpha\in[0, \, 1]$. Then
\begin{align}
    \notag
    P^{QM}_\uparrow(\hat n|\alpha\vec r) &= \frac{1}{2}(1+\alpha\vec r\cdot \hat n) 
    = \frac{1}{2}\left( 1 + 2\alpha \left( \frac{1}{2}\left(1 + \vec r\cdot \hat n\right) - \frac{1}{2}\right) \right) \\ \notag 
    &= \frac{1 -\alpha}{2} + \alpha P^{QM}_\uparrow(\hat n|\vec r).
\end{align}
This necessitates 
\begin{align}
    \notag
    &\int_{S^2}\dup A(\hat \lambda)\, p(\hat\lambda |\alpha \hat r) H(\hat n\cdot \hat \lambda) 
    \stackrel{!}{=} \frac{1-\alpha}{2} + \alpha \int_{S^2}\dup A(\hat \lambda)\, p(\hat\lambda | \hat r) H(\hat n\cdot \hat \lambda)  \\
    &= \int_{S^2}\dup A(\hat \lambda)\, \left(\alpha p(\hat\lambda | \hat r) +\frac{1-\alpha}{4\pi} \right) H(\hat n\cdot \hat \lambda).
\end{align}
We obtain the valid (and correctly normalized, non-negative) solutions (with~$\hat r = \vec r / \norm{\vec r}$)
\begin{align}
    \notag
    p(\hat \lambda | \vec r) &= p(\hat \lambda | \norm{\vec r}\hat r) = \norm{\vec r}p(\hat \lambda | \hat r) + \frac{1-\norm{\vec r}}{4\pi} 
    = \norm{\vec r} \frac{\hat r \cdot \hat \lambda}{\pi}H(\hat r\cdot \hat \lambda)+ \frac{1-\norm{\vec r}}{4\pi}\\
    &= \frac{\vec r\cdot \hat \lambda}{\pi}H(\vec r\cdot \hat \lambda) + \frac{1-\norm{\vec r}}{4\pi}.
\end{align}

\end{proof}

\end{document}